\documentclass{neutrinos19}

\def\PRD{{\em Phys. Rev.} D}

\def\be{\begin{equation}}
\def\ee{\end{equation}}
\def\bea{\begin{eqnarray}}
\def\eea{\end{eqnarray}}

\usepackage{tikz}
\newcommand*\Maru[1]{\tikz[baseline=(char.base)]{
            \node[shape=circle,draw,inner sep=0.5pt] (char) {#1};}}

\newcommand{\Photo}{\includegraphics[height=35mm]{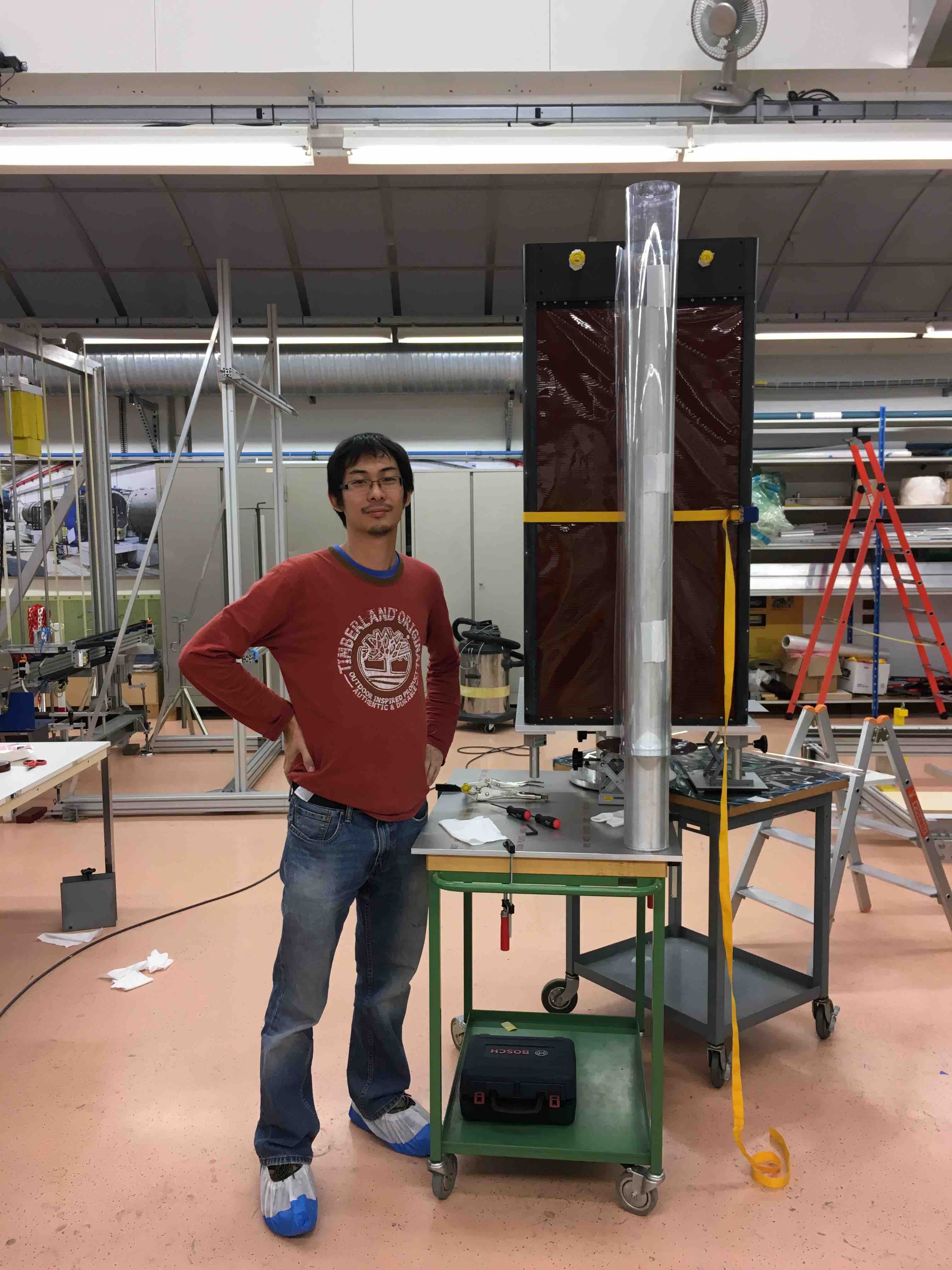}}

\begin{document}
\vspace*{4cm}
\title{Precise Determination of Neutrino Flux with Hadron Production Measurements}

\author{ Yoshikazu Nagai }

\address{Department of Physics, University of Colorado Boulder,\\
2000 Colorado Ave., Boulder, 80309, USA\vspace{3mm}}

\maketitle  \abstracts{
A precise prediction of the neutrino flux is a critical input for achieving the physics goals of accelerator-based neutrino experiments.
In modern experiments, neutrino beams are created from the decays of secondary hadrons produced in hadron-nucleus interactions.
Hadron production is the leading systematic uncertainty source on the neutrino flux prediction; therefore, its precise measurement is essential.
In this proceedings, recent results and ongoing analyses of hadron production measurements by the NA61/SHINE experiment, as well as prospect for achievable precision on the flux predictions of T2K and Fermilab-based long-baseline neutrino experiments, will be presented.
In addition, the necessity and prospects of further hadron production measurements by NA61/SHINE at CERN and the EMPHATIC experiment at Fermilab for the next generation neutrino experiments will be discussed.
}

\section{Introduction}

In modern accelerator-based neutrino experiments, high energy and intensity proton beams strike a long target which is typically made of graphite or beryllium. 
Secondary hadrons produced from the primary collisions can decay to neutrinos or can reinteract in the target or in beamline materials, such as in the aluminum magnetic focussing horns, potentially producing additional hadrons contributing to the neutrino flux.
These hadron production and reinteraction cross sections are critical inputs for the precise flux prediction in current and future accelerator-based neutrino experiments, however, these cross sections remain the leading systematic uncertainty on the neutrino flux prediction.

Neutrino flux predictions rely on hadronic interaction models, such as FLUKA (J-PARC/T2K) and GEANT4 FTFP\_BERT (NuMI beamline). 
However, model predictions vary from one prediction to the other.
For instance, it is found that the NuMI on-axis flux for the MINERvA experiment varies more than 40\% among GEANT4 hadronic interaction model predictions at the beam peak.
For precision neutrino measurements, it is important to constrain uncertainties coming from variation of model predictions using external hadron production data sets.

\section{Hadron Production Experiments}

Several hadron production measurements have been conducted in past, such as MIPP at Fermilab~\cite{Raja:2006pv}, HARP at CERN PS~\cite{Catanesi:2007ig}, and NA56/SPY at CERN SPS~\cite{Ambrosini:1999id}. 
Experiments currently on the hadron production measurement business are the NA61/SHINE~\footnote{The NA61/SPS Heavy Ion and Neutrino Experiment}$^,$\,\cite{Abgrall:2014xwa} and EMPHATIC~\footnote{Emulsion-based Measurement of Production of Hadron At a Test beam In Chicagoland} experiments.
These two experiments will be introduced in the following sections.

\begin{figure}
\centerline{\includegraphics[width=1.01\linewidth]{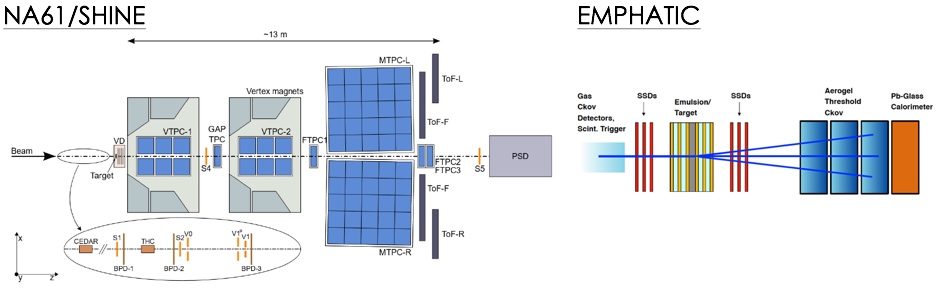}}
\caption[]{(Left) NA61/SHINE experiment. (Right) EMPHATIC experiment.}
\label{fig:NA61_EMPHATIC}
\end{figure}

\subsection{NA61/SHINE}\label{subsec:NA61}

NA61/SHINE is a fixed target experiment at the CERN Super Proton Synchrotron (SPS) which can receive secondary hadron beams between 13\,GeV/$c$ and 350\,GeV/$c$.
The NA61/SHINE detector is shown in Figure~\ref{fig:NA61_EMPHATIC} (Left).
It comprises  two superconducting magnets, five time projection chambers (TPC), a time of flight system (ToF), and a forward hadron calorimeter. 
In 2017, three new forward TPCs (FTPCs) have been built to fill the acceptance gap of the beam-forward region.
Upstream of the NA61/SHINE detector, three beam position detectors (BPDs) are located to reconstruct beam trajectories.
Scintillator counters and Cherenkov counters are used to trigger events and to select particle species.
It provides large acceptance for charged particles with good momentum and particle identification resolution using TPC and ToF.
Detector upgrade is ongoing to conduct further measurements after 2021, which will be discussed later.

\subsection{EMPHATIC}

EMPHATIC is a fixed target experiment at the Fermilab Test Beam Facility which can receive hadron beams between 0.2\,GeV/$c$ and 120\,GeV/$c$.
In 2018, EMPHATIC has completed test data taking using proton and kaon beams at 2, 10, 20, 30, and 120\,GeV/$c$ on various thin nuclear targets. 
The experimental setup used for the test data taking is shown in Figure~\ref{fig:NA61_EMPHATIC} (Right).
It comprises silicon and emulsion trackers followed by Aerogel Threshold Cherenkov detector and lead-glass calorimeter.
Its setup is dedicated for forward precision measurements, such as direct integrated cross section measurements of coherent elastic and quasi-elastic processes.
Analyses of test data sets are currently ongoing.
Future upgrade is under consideration, which will be discussed later.

\section{Thin Target Measurements}

Thin targets, which are typically a few \% of nuclear interaction length ($\lambda$), are used to study single interaction of hadrons.
Main objective is to measure integrated production and inelastic cross sections and differential cross sections of hadron production on nuclear targets.

\subsection{Production and Inelastic Cross Section Measurements}

First of all, it is worth noting that past and recent measurements do not always use the same terminology for production and inelastic  cross sections ($\sigma_\mathrm{prod}$ and $\sigma_\mathrm{inel}$). 
Through this proceedings, the inelastic and production cross sections are defined as follows:
\begin{equation}
\sigma_\mathrm{inel} = \sigma_\mathrm{total} - \sigma_\mathrm{el}, \label{eq:inel_xsec}\,\,
\sigma_\mathrm{prod} = \sigma_\mathrm{inel} - \sigma_\mathrm{qe} \label{eq:prod_xsec},
\label{eq:x-sec}
\end{equation}
where $\sigma_\mathrm{total}$, $\sigma_\mathrm{el}$, and $\sigma_\mathrm{qe}$ are total cross section, coherent elastic cross section, and quasi-elastic cross section.
The same definition is used for the tuning of T2K's beam flux simulation.
On the other hand, the tuning of NuMI beam flux simulation uses the term ``absorption cross section ($\sigma_\mathrm{absorption}$)'' for $\sigma_\mathrm{inel}$, while previous measurements sometimes refer to either $\sigma_\mathrm{prod}$ or $\sigma_\mathrm{inel}$ as $\sigma_\mathrm{absorption}$.
(for example, Carroll \textit{et al}.~\cite{Carroll:1978hc} used $\sigma_\mathrm{prod}$ as $\sigma_\mathrm{absorption}$, while Denisov \textit{et al}.~\cite{Denisov:1973zv} used $\sigma_\mathrm{inel}$ as $\sigma_\mathrm{absorption}$).

\begin{figure}
\begin{minipage}{0.5\linewidth}
\centerline{\includegraphics[width=1.1\linewidth]{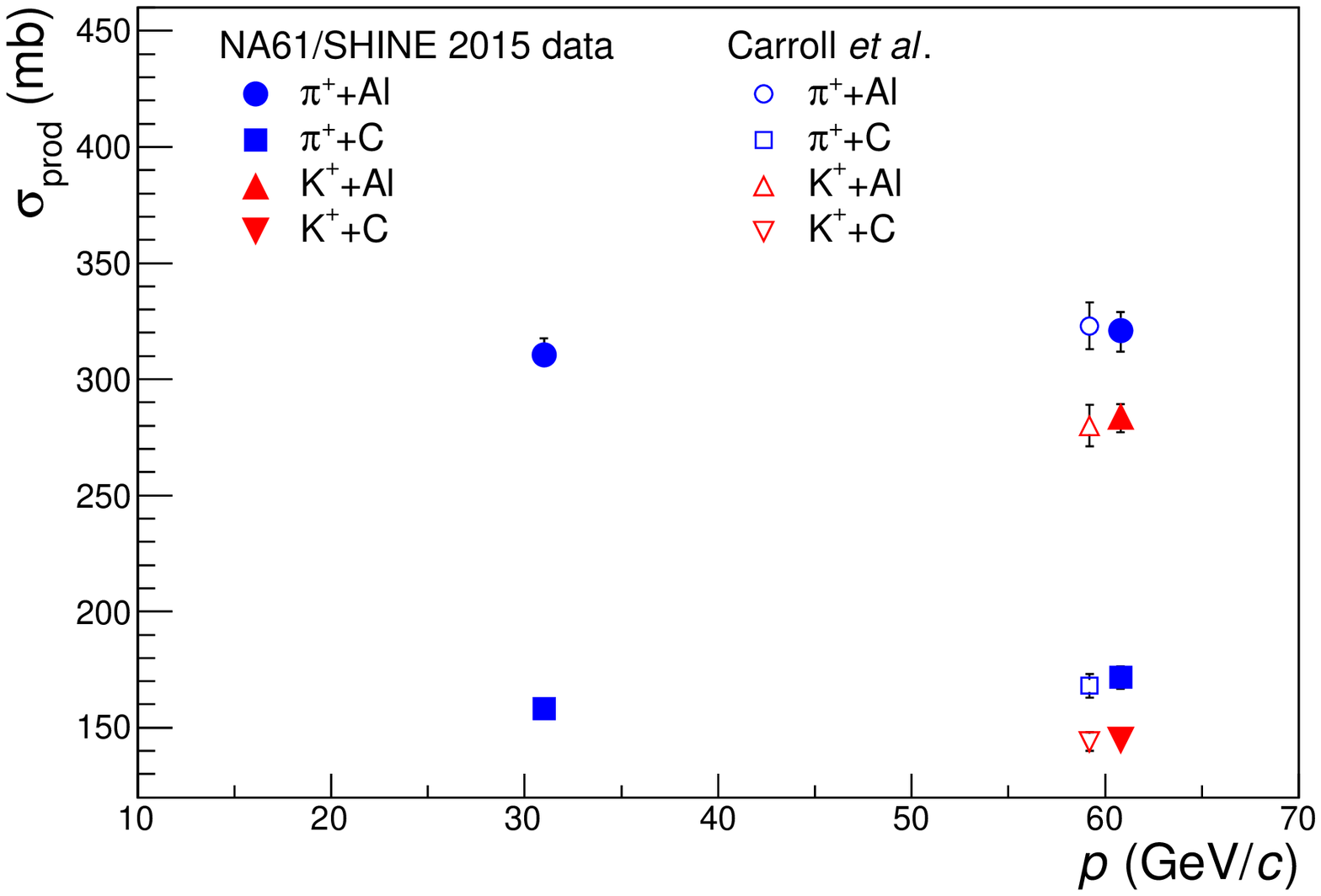}}
\end{minipage}
\hfill
\begin{minipage}{0.5\linewidth}
\centerline{\includegraphics[width=1.1\linewidth]{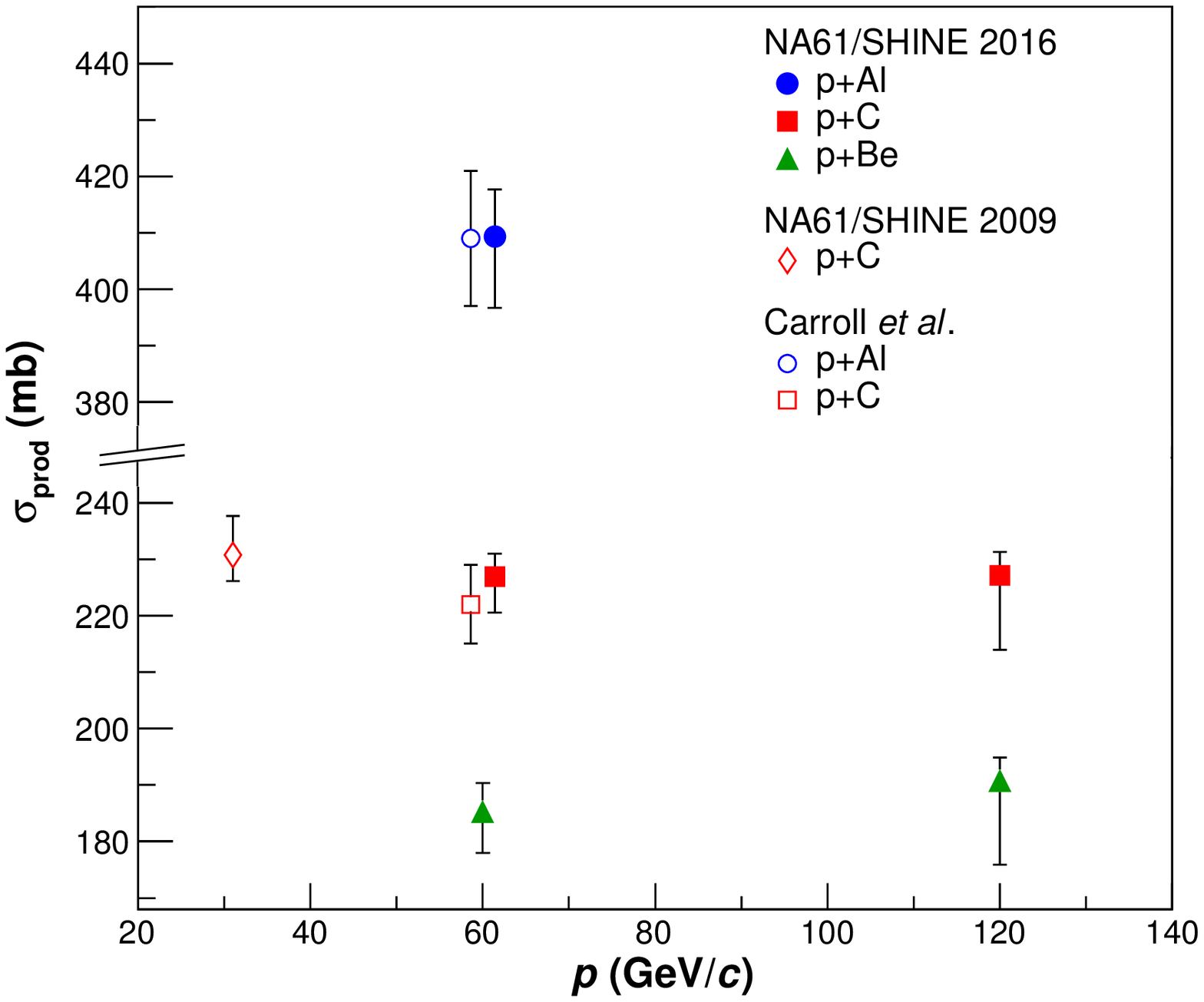}}
\end{minipage}
\hfill
\caption[]{Summary of production cross section measurements with NA61/SHINE. NA61/SHINE data has been compared to results by Carroll $\textit{et al}$~\cite{Carroll:1978hc}.  (Left) With positively charged pion and kaon beams. (Right) With proton beams.}
\label{fig:total-x-sec}
\end{figure}

NA61/SHINE has measured protons at 31\,GeV/$c$ on carbon targets (p+C @ 31\,GeV/$c$) for T2K~\cite{Abgrall:2015hmv}, 
$\pi^+/K^+$+C/Al @ 31 and 60\,GeV/$c$~\cite{Aduszkiewicz:2018uts,Aduszkiewicz:2019hhe}, 
and p+C/Be/Al @ 60 and 120\,GeV/$c$~\cite{Aduszkiewicz:2019xna}.
Summary of $\sigma_\mathrm{prod}$ measurements is shown in Figure~\ref{fig:total-x-sec}.

Measured cross sections are then used to constrain interaction rate in MC simulation by calculating the weight:
\begin{equation}
W = \frac{\sigma_\mathrm{NA61\,data}}{\sigma_\mathrm{MC}}e^{-x(\sigma_\mathrm{NA61\,data} - \sigma_\mathrm{MC})\rho},
\label{eq:x-sec}
\end{equation}
where $\sigma_\mathrm{NA61\,data}$ is measured cross section by NA61/SHINE, $\sigma_\mathrm{MC}$ is a cross section prediction by a hadronic interaction model, $x$ is travel distance of particle, and $\rho$ is material density, respectively.
These weights are applied to every particles. 
Moreover, these measurements are also important to determine normalization of differential cross section measurements.

\subsection{Differential  Cross Section Measurements}

\begin{figure}
\centerline{\includegraphics[width=1.0\linewidth]{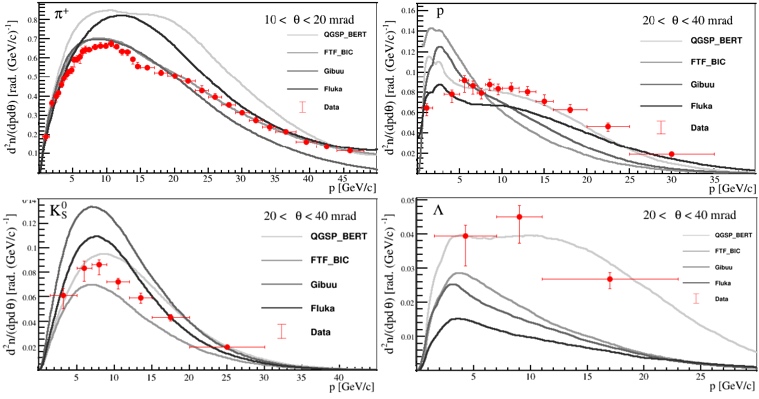}}
\caption[]{Measurement of differential multiplicity from $\pi^+$+C @ 60 GeV/$c$ interactions. NA61/SHINE data has been compared with several model predictions (GEANT4 QGSP\_BERT, GEANT4 FTF\_BIC, Gibuu, and FLUKA).}
\label{fig:diff-spectra}
\end{figure}

NA61/SHINE has measured differential production multiplicities of charged hadrons ($\pi^\pm$, $K^\pm$, $p$) and neutral hadrons ($K^0_\mathrm{S}$, $\Lambda$, $\bar{\Lambda}$) in slices of momentum ($p$) and production angle ($\theta$) of hadrons ($d^2n/dpd\theta$).
These can be transformed into differential cross sections by normalizing by total cross sections ($d^2\sigma/dpd\theta$).
Measurements on p+C @ 31\,GeV/$c$ were made for T2K~\cite{Abgrall:2015hmv} and recently measurements on $\pi^+$+C/Be @ 60 GeV/$c$ have been made for the Fermilab beamlines~\cite{Aduszkiewicz:2019hhe}.
An example of multiplicity spectra from $\pi^+$+C @ 60 GeV/$c$ is shown in Figure~\ref{fig:diff-spectra}.

Measured differential production multiplicities are then used to rescale MC predictions by calculating the weight for a given kinematic bin in $p$ and $\theta$: 
\begin{equation}
W_i(p, \theta) = \frac{N_i(p,\theta)_\mathrm{NA61\,data}}{N_i(p,\theta)_\mathrm{MC}}, 
\label{eq:x-sec}
\end{equation}
where $N_i(p,\theta)_\mathrm{NA61\,data}$ is number of produced hadrons measured by NA61/SHINE, $N_i(p,\theta)_\mathrm{MC}$ is number of produced hadrons predicted by a hadronic interaction model, and $i$ represents one of particle species among $\pi^\pm, K^\pm$, proton, $K^0_\mathrm{S}, \Lambda$, or $\bar{\Lambda}$.

\section{Replica Target Measurements}

Replica targets of long-baseline experiments, often also referred as thick targets, are used to study hadrons exiting from the surface of replica targets and beam attenuation inside replica targets.
NA61/SHINE has collected data with a T2K replica target (90\,cm, $\lambda\!\sim\!1.9$) and a NOvA replica target (130\,cm, $\lambda\!\sim\!2.5$).
Main objective is to measure differential hadron production yields ($d^3\sigma/dpd\theta dz$) and beam survival probability ($P_\mathrm{survival} = e^{-Ln\sigma_\mathrm{prod}}$, $L$: length of target, $n$: number of atoms per unit volume).  

NA61/SHINE completed data collection using the T2K replica target and data analysis is being finalized. 
NA61/SHINE has also collected data using the NOvA replica target in 2018.

\subsection{T2K Replica Target Measurements}
\begin{figure}
\centerline{\includegraphics[width=1.0\linewidth]{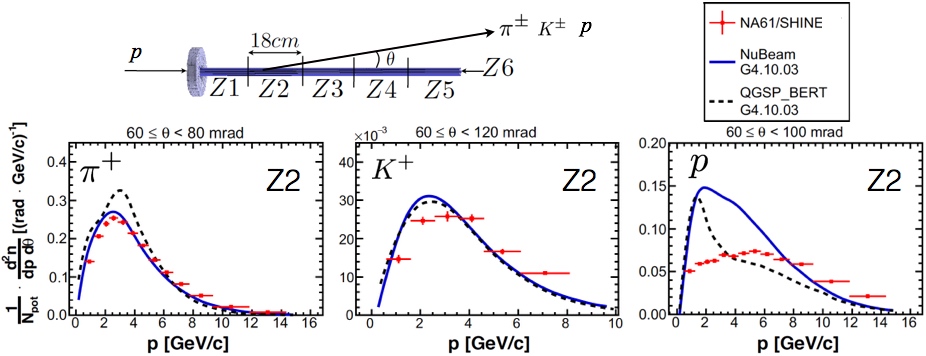}}
\caption[]{Measurement of differential hadron yields. NA61/SHINE data has been compared with two GEANT4 models, NuBeam and QGSP\_BERT.}
\label{fig:t2k-multiplicity}
\end{figure}

NA61/SHINE has recently measured $\pi^\pm$, $K^\pm$, and proton yields from the surface of the T2K replica target with a 31\,GeV/$c$ proton beam~\cite{Berns:2018tap}. 
This analysis used higher data statistics compared to former NA61/SHINE replica target measurements, about 10 million protons on target,
and measured $K^\pm$ and proton spectra for the first time, in addition to $\pi^\pm$ spectra.
An example of multiplicity spectra normalized to number of protons on target is shown in Figure~\ref{fig:t2k-multiplicity}.

\begin{figure}
\centerline{\includegraphics[width=0.68\linewidth]{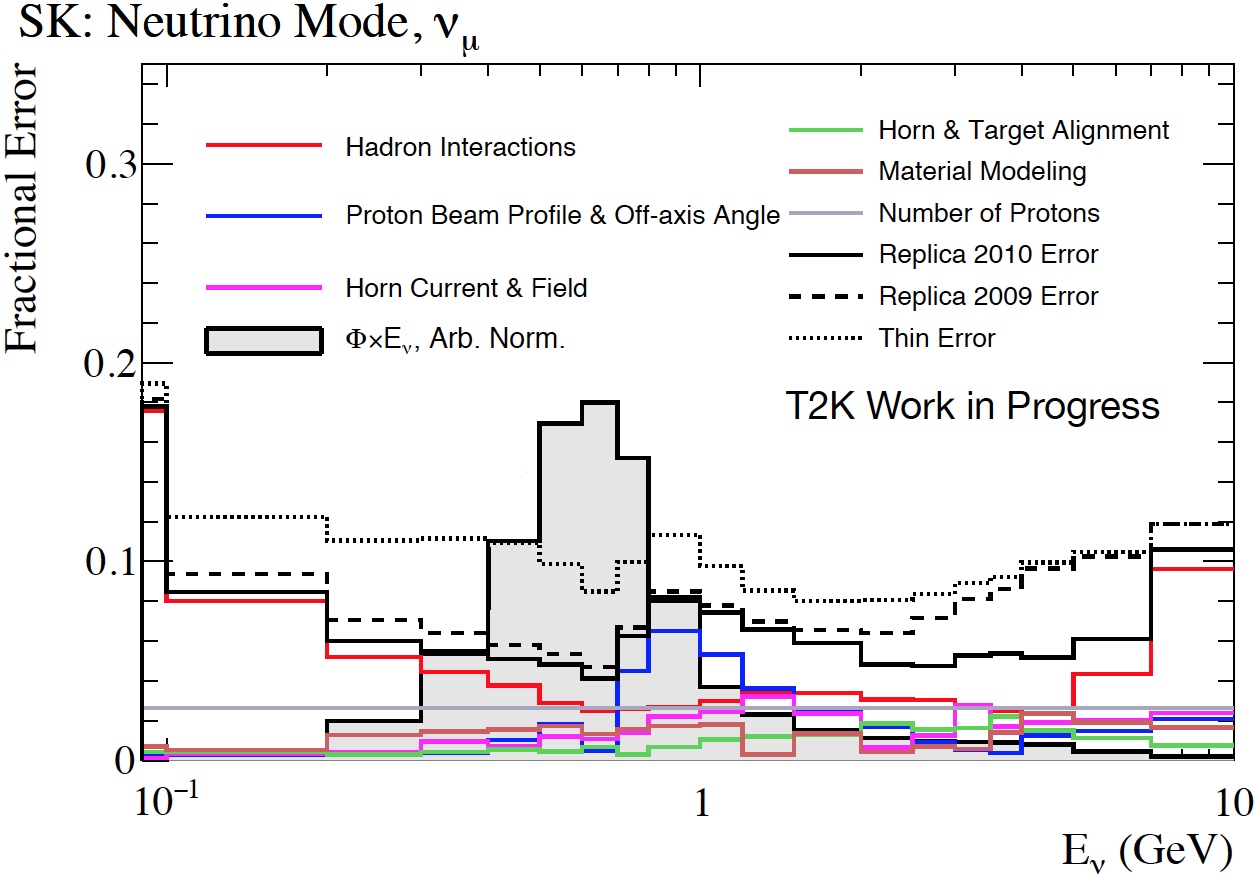}}
\caption[]{Tuned T2K flux uncertainty with NA61/SHINE's thin target measurement (black dotted line), with NA61/SHINE's previous replica target measurement (black dashed line), and with the latest replica target measurement (black solid line).}
\label{fig:T2KtunedFluxUnc}
\end{figure}

Measured differential production multiplicities are then used to rescale MC predictions for each hadrons exiting target surface by calculating the weight for a given kinematic bin in $p$, $\theta$, and position along with the beam direction ($z$):
\begin{equation}
W_i(p, \theta, z) = \frac{N_i(p,\theta,z)_\mathrm{NA61\,data}}{N_i(p,\theta,z)_\mathrm{MC}}, 
\label{eq:t2kReplicaTuning}
\end{equation}
where $N_i(p,\theta,z)_\mathrm{NA61\,data}$ is number of produced hadrons exiting target surface measured by NA61/SHINE, $N_i(p,\theta,z)_\mathrm{MC}$ is number of produced hadrons predicted by a hadronic interaction model, and $i$ represents one of particle species among $\pi^\pm, K^\pm$, and proton.
Together with the high statistic NA61/SHINE replica target measurement,
T2K expects to achieve flux uncertainty down to $\sim$4\% level at the beam peak as shown in Figure~\ref{fig:T2KtunedFluxUnc}.

\begin{figure}
\centerline{\includegraphics[width=1.03\linewidth]{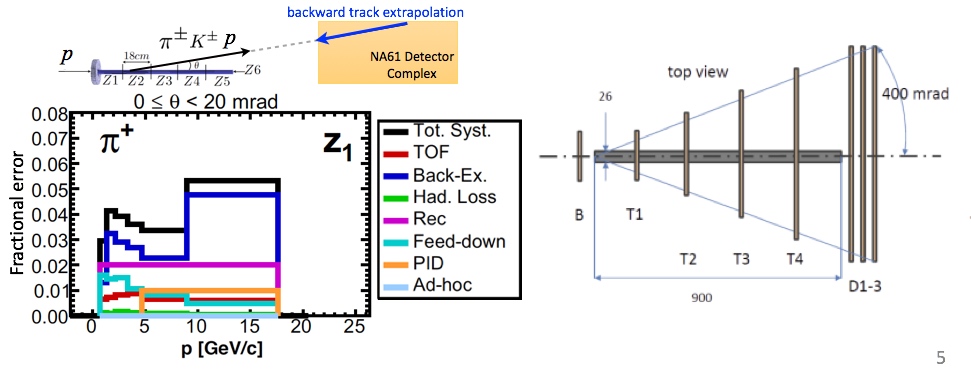}}
\caption[]{(Left) Uncertainty on T2K replica measurement at the most upstream $z$ bin. (Right) A conceptual drawing of a new tracking detector surrounding the long target.}
\label{fig:TrackExtrapolateUnc}
\end{figure}

T2K replica target measurement has been successfully made but confirmed the necessity to improve track position resolution.
Particularly it is relevant for the upstream $z$ bins due to long distance backward track extrapolation, as shown in Figure~\ref{fig:TrackExtrapolateUnc} (Left).
This backward tracking becomes a dominant source of the systematic uncertainty. 
Building an additional tracking detector surrounding the long target is under discussion to reduce this uncertainty.
A conceptual drawing of a new tracking detector is shown in Figure~\ref{fig:TrackExtrapolateUnc} (Right).

In addition, measurement of beam survival probability is ongoing to study beam interaction rate in the target.
NA61/SHINE has recorded a special data set with the maximal magnetic field, resulting 31\,GeV/$c$ proton beam passing through the replica target bends into a TPC and is able to reconstruct beam trajectory in the detector.
This will be important information to understand discrepancy which we observed between thin and thick flux tuning, which will be discussed later.

\subsection{NOvA Replica Target Measurements}

\begin{figure}
\centerline{\includegraphics[width=0.85\linewidth]{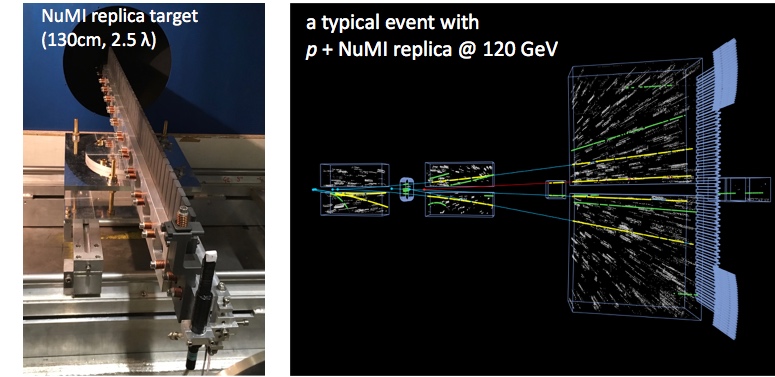}}
\caption[]{(Left) NOvA replica target installed to the NA61/SHINE facility. (Right) A typical event with the NOvA replica target.}
\label{fig:NOvAreplica}
\end{figure}

In 2018, NA61/SHINE has collected data with a 120\,GeV/$c$ proton beam on target on a NOvA replica target.
Experimental setup of the NOvA target and a typical event are shown in Figure~\ref{fig:NOvAreplica}.
Currently, calibration of the high statistics data set (about 18 million protons on target) is ongoing and results will be expected near future, which will improve the flux uncertainties of NuMI beamline experiments, NOvA and MINERvA.
It is worth noting that NOvA replica target may require quad differential yields measurement ($d^4\sigma/dpd\theta d\phi dz$) because the NOvA target has rectangular shape and it is not symmetric in $\phi$ coordinate \footnote{T2K target has a cylindrical shape, therefore, it is symmetric in $\phi$ coordinate.}.

\subsection{Thin vs Replica Target Tuning}
\begin{figure}
\centerline{\includegraphics[width=1.0\linewidth]{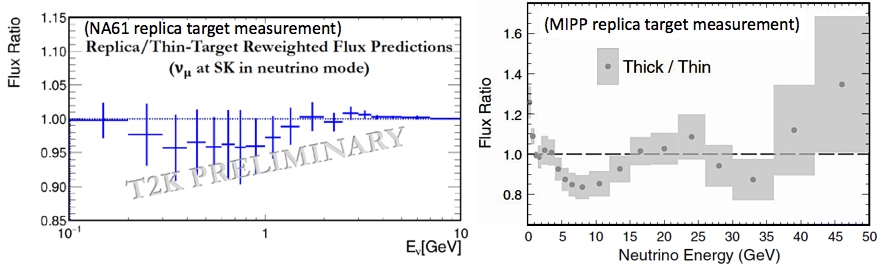}}
\caption[]{Thin and replica flux tuning comparison. (Left) T2K tuning comparison using NA61/SHINE data. (Right) NuMI/MINERvA comparison using MIPP data.}
\label{fig:thin_vs_replica}
\end{figure}

Consistency checks of thin-based flux tuning and replica-based flux tuning are performed.
In the case of T2K, the ratio between the thin and replica target flux predictions agree within the error band (Figure~\ref{fig:thin_vs_replica} Left), however, 
the replica target $\nu_\mu$ flux prediction tends to lower than the thin target prediction around the flux peak, 600\,MeV~\cite{Vladisavljevic:2018prd}.
A more clear discrepancy was observed in the case of the NuMI target (the Low-Energy configuration) for the MINERvA experiment using replica target hadron production measurements conducted by the MIPP experiment, as shown in Figure~\ref{fig:thin_vs_replica} (Right)~\cite{Aliaga:2016oaz}. 

The cause of this discrepancy has yet to be understood, however, there are several possibilities.
One is the beam interaction rate in the target.
The T2K target is shorter than the NuMI target, and it has less interaction rate for incoming protons.
To investigate the possible sources of discrepancy, further studies with high precision thin and replica target measurements are necessary. 
For thin target tunings, current production and inelastic cross section measurements have relatively large uncertainties, and 
higher precision production and inelastic cross section measurements are desirable together with direct measurements of elastic and quasi-elastic components.
For replica target tunings, measurement of beam attenuation will give a handle to investigate interaction rate in the target.

The ``Wiggle'' shape of the NuMI/MINERvA flux prediction was not seen in the T2K flux prediction, 
and it may relate to beamline intrinsic effects, such as difference on beam optics configuration, focusing parameters, or beam monitor system.

\section{Future Prospect}

The next generation experiments, Hyper-K and DUNE, will need to achieve total systematic uncertainty below 4\% including flux uncertainty.
The goal for the flux uncertainty is the 2\% level, which is an unprecedented precision.
Therefore, we need to collect all the necessary data to constrain the dominant component of the flux uncertainty, hadron production.
The NA61/SHINE and EMPHATIC experiments will be leading experiments to achieve this ambitious goal.

The NA61/SHINE collaboration plans to extend their physics program for 2021-2024 including significant facility upgrades.
For neutrino physics, next upgrades are particularly relevant: DAQ upgrade with $\sim$1\,kHz TPC readout (current readout rate is $\sim$100\,Hz), new ToF walls with mRPC technology with better timing resolution, construction of a low-momentum tertiary beamline (1-10\,GeV/$c$), and new replica target tracking detector.
With these upgrades, NA61/SHINE will conduct further thin target measurements (e.g. low-momentum 1-5\,GeV/$c$ $\pi^\pm$ interactions for T2K, 30-60\,GeV/$c$ $\pi^\pm$/$K^\pm$ interactions for DUNE) and new replica target measurements with Hyper-K and DUNE targets.

Upgrades of the EMPHATIC detector are under consideration with larger acceptance and a permanent magnet.
It comprises silicon trackers, Aerogel RICH counters, RPC ToF counters, and lead glass calorimeter.
With these upgrades, EMPHATIC will conduct precise thin target measurements utilizing various beam momentum.

It is worth noting that these future measurements by two experiments are complementary rather than competitive in following reasons:
\Maru{1} Replica target measurements will be only possible by the NA61/SHINE facility. 
\Maru{2} Two thin target measurements with different detector technology will allow to cross-check and combine results.
\Maru{3} EMPHATIC's silicon-based tracker will allow precise forward measurements of direct elastic and quasi-elastic processes.

\section{Summary}

Precise hadron production measurements are essential to reduce the leading systematic uncertainty on the neutrino flux prediction. 
Thin and replica target measurements by the NA61/SHINE experiment will reduce the flux uncertainty down to the 4\% level for T2K 
and rich hadron production data has been collected and being analyzed for the Fermilab NuMI and LBNF beamlines.
A dedicated forward measurement has started to understand integrated coherent elastic and quasi-elastic cross sections more precisely by the EMPHATIC experiment.

Further precise hadron production measurement is necessary for future long-baseline neutrino experiments, Hyper-K and DUNE.
Significant upgrades of experimental facility are planned and ongoing by the NA61/SHINE and EMPHATIC collaborations.
These upgrades will allow new measurements with thin and replica targets to satisfy the requirement of next generation accelerator-based neutrino experiments.

\section*{Acknowledgments}

This work was supported by the U.S. Department of Energy.

\end{document}